# A Taxonomy for Congestion Control Algorithms in Vehicular Ad Hoc Networks

Mohammad Reza Jabbarpour Sattari[1], Rafidah Md Noor[2], Hassan Keshavarz[3]
Faculty of Computer Science and Information Technology, University of Malaya, Kuala Lumpur, 50603 Malaysia
[1]Reza.jabbarpour@ieee.org, [2]fidah@um.edu.my, [3]keshavarz_hassan@ieee.org

***Abstract*—One of the main criteria in Vehicular Ad hoc Networks (VANETs) that has attracted the researchers' consideration is congestion control. Accordingly, many algorithms have been proposed to alleviate the congestion problem, although it is hard to find an appropriate algorithm for applications and safety messages among them. Safety messages encompass beacons and event-driven messages. Delay and reliability are essential requirements for event-driven messages. In crowded networks where beacon messages are broadcasted at a high number of frequencies by many vehicles, the Control Channel (CCH), which used for beacons sending, will be easily congested. On the other hand, to guarantee the reliability and timely delivery of event-driven messages, having a congestion free control channel is a necessity. Thus, consideration of this study is given to find a solution for the congestion problem in VANETs by taking a comprehensive look at the existent congestion control algorithms. In addition, the taxonomy for congestion control algorithms in VANETs is presented based on three classes, namely, proactive, reactive and hybrid. Finally, we have found the criteria in which fulfill prerequisite of a good congestion control algorithm.**

## I. Introduction

VANET belongs to wireless communication networks area. The frequency spectrum for VANET's wireless communication is allocated by The Federal Communication Commission (FCC). Then the Dedicated Short Range Communications (DSRC) Service was established by the Commission in 2003. The DSRC is a communication service which is utilized for public and private safety and uses the frequency range of 5.850-5.925 GHz [1]. The DSRC was designed into multi-channel system. The DSRC spectrum is divided into seven channels by the FCC so that each of them has 10 MHz bandwidth. Six of them were identified as Service Channels (SCH), and one of them is identified as the Control Channel (CCH), as shown in Fig. 1. The CCH channel is used for safety messages, while SCH channels are used for non-safety as well as WAVE-mode messages or services [4], [5].

VANET's aim is to increase the safety of road users and comfort of passengers. The safety messages can be categorized into two categories; beacon and event-driven messages. Beacon messages send periodically by vehicles to inform their condition such as position, direction and speed to their neighbor vehicles. The beacon messages are used by the neighboring vehicles (nodes) to be aware of their environment as well as preventing potential dangers [6], [7]. The event-driven safety messages are generated when an abnormal condition or an imminent danger is detected and are disseminated within a certain range with higher priority [7]. The event-driven safety messages should be delivered to neighboring node by high reliability and limited time. A single delayed or lost message could result in loss of life [7].

VANET is a special type of MANET, in which the vehicles act as nodes. Unlike MANET, vehicles move on predefined roads, vehicles velocity depends on the speed signs and in addition these vehicles also have to follow traffic signs and signals [8].

There are many problems in VANET that should be solved in order to provide reliable services such as routing, security, quality of service. Dynamic topology, lack of central coordination, error prone shared radio channel, hidden terminal problem, limited resource availability and insecure medium are challenges which make VANETs inefficient to support Quality of Service (QoS). There are many techniques proposed for improving the QoS in VANETs where one of them is congestion control. Thus, previous works on congestion control algorithms for VANETs are discussed in this paper and taxonomy of congestion control is presented using a classification technique.

The rest of the paper includes 4 Sections and structured as follow: Section 2 presents congestion control classes which are provided into three categories of proactive, reactive and hybrid. Section 3 discusses about various congestion control algorithms in VANETs based on predefined classes in Section 2. Taxonomy of congestion control algorithms in VANETs is illustrated in Section 4. Section 5 concludes the paper with outlooks on the future work.

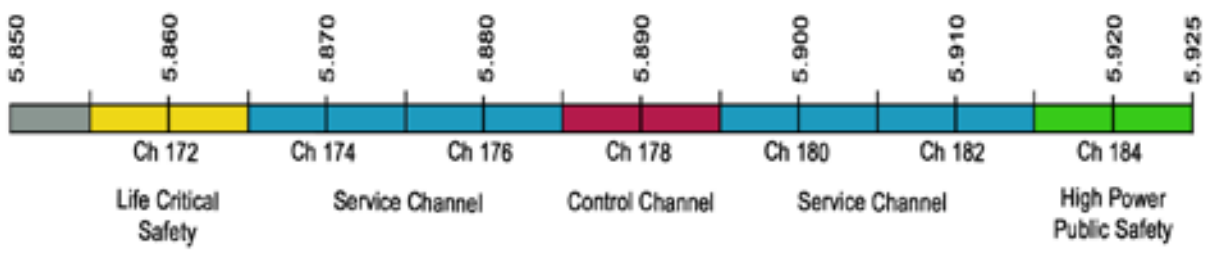

Figure 1. DSRC channels

## II. Congestion Control Classes for Vanets

The major classification criterion considers the information base from which congestion control mechanisms derive their decision to adjust the transmission parameters. The first class is reactive congestion control which uses first-order information about the channel congestion status to decide whether and how an action should be undertaken. Because of their nature, actions to lessen channel load are undertaken only after a congested situation has been detected.

The second class is proactive congestion control which uses models, based on information such as number of nodes in the vicinity and data generation patterns, try to estimate transmission parameters which will not lead to congested channel conditions, while at the same time providing the desired application-level performance. In particular, such mechanisms typically employ a system model to estimate the channel load under a given set of transmission parameters, and make use of optimization algorithms to determine the maximum transmit power and/or rate setting that will adhere to a maximum congestion limit. [9].

The third class is hybrid congestion control which combined the advantages of both proactive and reactive approaches. The relative advantages and disadvantages of proactive vs. reactive approaches are discussed here.

Given their ability to prevent congestion, proactive approaches are very appealing for vehicular environments, where radio communications are primarily used for safety applications, whose performance would be seriously threatened by congested channel conditions. However, proactive approaches come with two major drawbacks.

First, in order to estimate the expected load generated by neighboring vehicles, such approaches require a communication model that maps individual transmission power levels to deterministic carrier sense ranges. However, this mapping is reasonable only as long as it reflects the average propagation conditions of the wireless channel. Thus, propagation conditions should be either dynamically estimated as the vehicle moves, which is very difficult to do in a practical scenario, or they should be statistically estimated to build specific profiles for different environments, e.g., urban and highway.

Second major drawback of proactive approaches is the need to carefully estimate the amount of generated application-layer traffic in a certain period of time. Although in some cases this is indeed possible (e.g., in the case of applications built on top of periodic beacon exchange), accurate application-layer traffic estimation is a challenging task in general.

Reactive approaches, which do not suffer of the drawbacks that accompany proactive mechanisms, nonetheless have the notable disadvantage of undertaking control actions only after a congested channel condition has been detected. Considering that some time is needed to recover from a congested channel situation, this means that reactive approaches expose safety-related applications to the risk of not being able to fulfill their design goal, due to the poor (temporary) performance of the underlying radio channel. Another disadvantage of reactive approaches is that the important design goals such as fairness and packet prioritization are more difficulties to achieve than in a proactive approach. We remark that fairness is the important in vehicular networks to ensure that all vehicles in the network have similar opportunities to communicating with nearby nodes. In fact, if congestion control were to be obtained by sacrificing, say, a specific node in the network is forced to set its transmission power to a very low value, this node would not have a chance to communicate with nodes in its surrounding, impairing application-level performance..

Most importantly, in safety-related applications, every vehicle in the network should be able to receive latest information about the status of the other vehicles in the surrounding, as well as to communicate its own status to the surrounding vehicles. Hence, fairness becomes a major design goal in safety-related applications. As for prioritization, providing a strict prioritization of different classes of packets is an important requirement for vehicular networking, which is partly addressed in the drafted IEEE 802.11p standard by adopting the Enhanced Distributed Channel Access (EDCA) mechanism defined within IEEE 802.11e [9].

## III. Congestion Control Approaches for Vanets

In this section, proposed congestion control approaches for VANETs will be reviewed based on above mentioned classification.

### A. *Proactive Congestion Control Approaches for Vanets*

According to the terminology which already defined, the Vehicle-To-Vehicle communication protocol approach [10] for Cooperative Collision Warning (VCCW), belongs to the class of proactive approaches, and acts on packet generation rate to prevent congestion. Yet, the approach of [10] is mostly an open-loop controller, since the multiplicative rate decreasing algorithm that is used to tune the packet generation rate is based only on predicted performance based on suitable models of the communication channel. On the other hand, a form of primary feedback (e.g., reception of redundant transmissions from following vehicles) is used in the decision rules to freeze emergency message transmission.

L. Wischhof and H. Rohling [11] proposed hop-by-hop proactive congestion control approach for VANET's. An innovative method related to congestion control base on utility function and also packet forwarding is presented by them in VANET's. This algorithm employs an application-specific utility function and in a transparent way, encrypts the quantitative utility information in each data packet which is transmitted, for all users within a local environment. Then, a regionalized algorithm computes the "Average Utility Value" of every single node according to its data packets utility value and later a part of the accessible data rate will be allocated in proportion to its relative priority. In order to evaluate the performance, the proposed decentralized Utility-Based Packet Forwarding and Congestion Control (UBPFCC) is initiated on top of the IEEE 802.11 MAC protocol.

Based on network simulations, which also contain the models of vehicular mobility, the UBPFCC prevents some nodes starvation within the network as well as demonstrates a considerable increase in the effectiveness level of dissemination and fairness of data [11]. This approach requires context exchange between neighbor nodes, which generates a communication overhead [2]. Moreover, since for the calculation of the message utility metric, the road is required to be sectioned in this approach, its being directly used in context of safety applications is not feasible [12]. Another problem that can be mentioned regarding this algorithm would be evaluating the message priority based on utility and packet size, which is reducing the disseminating performance of event-driven safety messages [6].

M.S. BOUASSIDA and M. SHAWKY [2] proposed another proactive congestion control approach, Dynamic Priority-Based Scheduling (DPBS), which considered above drawbacks in hop-by-hop approaches. Their application-layer approach is based on the concept of dynamic priority-based scheduling and is designed to guarantee communication architecture within VANET which is reliable and safe. In this model, packets with high priority are transmitted first and with no delay. Meanwhile, packets with medium or low priority are rescheduled. Scheduling process of the messages is done based on priorities and is calculated as a function of the utility of the related messages, the sender application and the neighborhood context. Dynamic priority assignment, messages scheduling and messages transmission are the major steps in this approach.

The messages generators give a priority to the packets in priority assignment, when they are created. There are two fields to determine priority of a packet: static field which is taken from the type of application and next dynamic field which is derived from VANET's specific context and is assigned by the module of congestion control. The overall priority indicator will be determined by the combination of dynamic and the static fields.

In scheduling of messages, each node schedules the messages based on their priorities. Scheduling process is divided into two types: the static process involves affecting messages based on each message's priority and directing them into the proper queues of communication channel and dynamic scheduling process which happens periodically, observes the message queues and calculates the total priority indicator of each single message. Then the messages are reordered based on the newly calculated priority by the rescheduling process.

Whenever the corresponding channel is free, the message with the highest priority will be sent, in messages transmission process. Yet, the high priority packets which are sent through control channel are preemptive as compared to those which were sent through the service channel. Even if the corresponding channels are not busy, transmission of the packets with lower priority will be stopped, so as to send packets with high priority with minimum delays. Additionally, whenever a higher priority message (compared to first messages in its queues that it will send) is received, it freezes the sending process [2]. The result showed that the delay of event-driven safety message in this approach is 50 ms in the worst scenario. This result is critical because pre-crash sensing safety application message needs to be disseminated to adjacent nodes within 20 ms [6].

Researchers in [2], [11] proposed the utilization of packets as an important part in their proposed methods. However, researchers in [12], [13] which will be discussed below, considered transmission power control in their congestion control approaches.

Researches on above mentioned approaches have been increased when researchers proposed Broadcast Reception Rates and Effects of Priority Access (BRR-EPA) algorithm in [13]. The purpose of transmit power control is to optimize energy consumption as well as point-to-point communications connectivity. Usually, a higher data rate needs a higher power of transmit from a sender; as a result it may generate a higher interference with neighboring nodes. By using dynamic power of transmission for congestion control, the level of channel usage will be limited; however, it can be used for dynamic reservation of a part of available bandwidth for the safety application. The main idea is to keep control of the low priority messages transmit power and maintain the highest priority traffic transmit power in its maximum level [13], [14].

J. Zang, et al. in [12] also proposed proactive Cross-layer Congestion Control (Cross-layer CC) via dynamic transmit power control. The performance of Emergency Electronic Brake Light with Forwarding (EEBL-F) application was studied by them as an example of safety measures in congested situations and consequently, they have recommended a cross-layer congestion control architecture for utilizing in VANET, which shown in Fig. 2.

They focused on Medium Access Control (MAC) layer, and presented two congestion detection methods, namely event-driven and measurement-based detection, as well as two congestion control approaches which are congestion

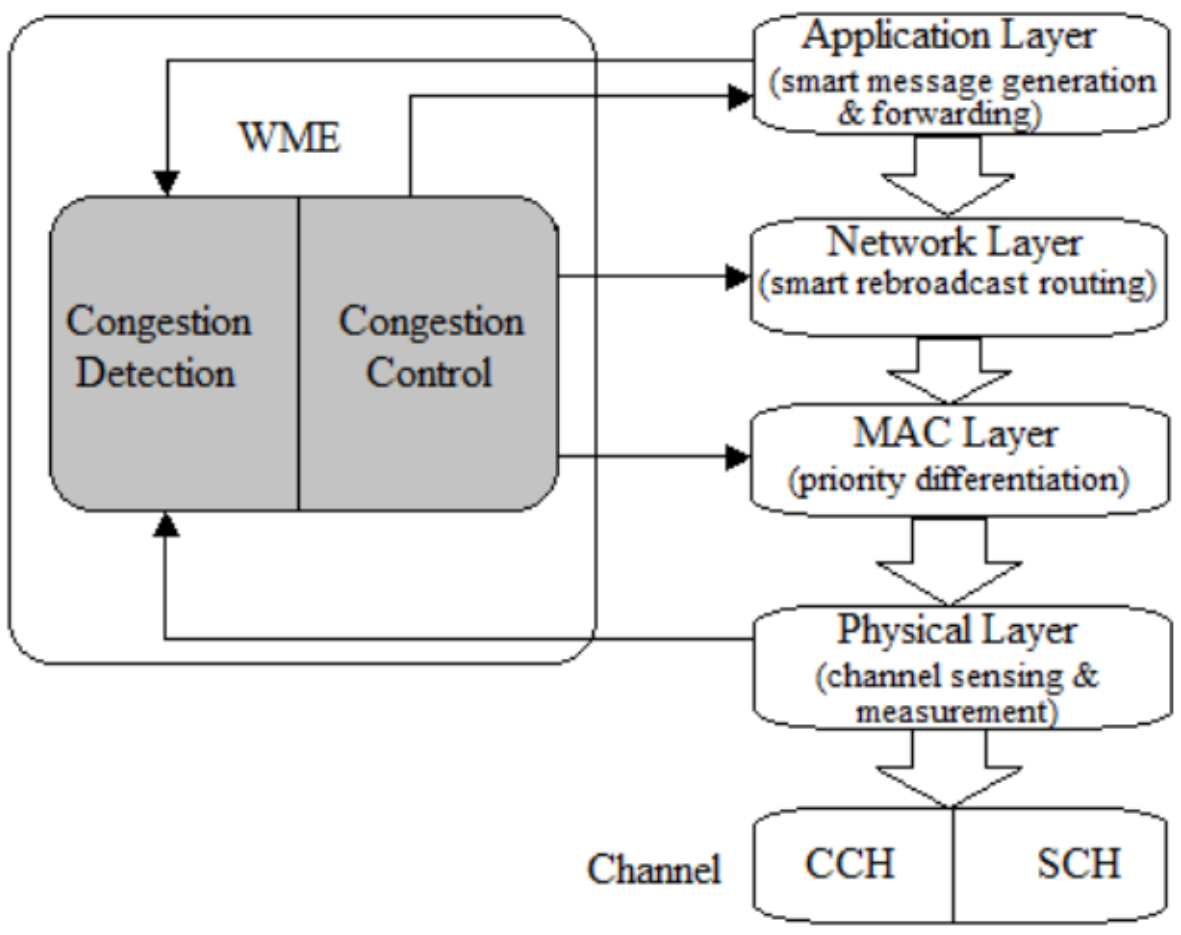

Figure 2. Cross-Layer Congestion Control Architecture

control via MAC queue manipulation and congestion control via dynamic transmit power control.

In event-driven detection method, each node applies the brute force queue freezing for all MAC transmission queues except for the safety queue with the highest priority. For example, when a node detects event-driven safety message either generated at its own application layer or received from another device, it will immediately launch the congestion control to guarantee the delivering of event-driven safety message [12], this method also been used in [2].

In the measurement-based detection method, each device periodically senses the channel based on the predefined thresholds. Different researchers applied different sensing threshold such as channel usage level in [12], number messages queue in [2] and channel occupancy time in [15].

The wireless networks performance is, considerably, affected by the predefined threshold which uses monitoring and detection of congestion in communication channels. The predefined threshold will be computes by above mentioned metrics. For example in [8], the level of channel usage is applied as threshold. By using this strategy, each device senses the usage level of channel periodically, and detects the congestion whenever the measured channel usage level exceeds the predefined threshold.

In congestion control via MAC queue manipulation, the main idea is to provide the safety message absolute priority over other traffic via manipulating the MAC transmission queues of lower prioritized traffics, or to dynamically reserve a fraction of bandwidth for the highest priority traffic with adaptive QoS parameters.

The transmit power sat for all packet types and investigate the control of transmit power effects on the congestion problem in VANET by J. Zang, et al in [12]. By increasing the level of transmit power, the IEEE 802.11p physical layers (PHY) can provide communications within a distance from 100m to 1km in vehicular environments. This algorithm need places equipped with Road Site Units (RSUs). Moreover, they focused only on the performance of the EEBL-F safety application. This congestion control algorithm should be testing on other event-driven safety applications such as pre-crash sensing and lane change warning [3].

Another proactive congestion control approach, Application-Based Congestion Control, proposed in [16]. An innovative congestion control policy was proposed for vehicular ad-hoc networks by M. Sepulcre, et al. Based on this policy, the communication parameters in each vehicle are modified according to their individual utilization necessities. Based on the policies in other approaches, transmission resources are normally allocated according to performance metrics at system-level. Unlike these approaches, the newly proposed technique independently satisfies the target application performance of every single vehicle, while generally it minimizes the load in the channel in order to avoid congestion in the channel. The outcomes show that various arrangements of transmission power and packet rates will be able to satisfy the utilization requirements. This method has been appraised by taking the lane change assistance application into consideration. Nevertheless, to competently allocate the accessible bandwidth, according to the experienced channel load, this method can be utilized as the foundation of advanced contextual congestion control policies and congestion control protocols. Since every single application has different requirements and is concurrently run by the same vehicle, this method cannot fulfill multiple applications, since particular applications require high packet rates and some have a constant minimum distance at which the messages are received. Deployment of these methods at VANET is challenging tasks, since these necessities should be safely combined for the future of [16].

Researchers in [17] proposed another proactive approach which is a fully distributed and localized algorithm called D-FPAV. They use transmit power control to achieve congestion control, fairness and prioritization in VANETs. They defined MaxBeaconingLoad (MBL) threshold for obtaining their first and second goals. Besides, they used “water-filling” approach [18] for getting their last goal. This algorithm has two weaknesses which are communication overhead can grow to 40% - compared to the actual application-layer data [9] and bandwidth wastage because of fixed MBL assignment.

### B. *Reactive Congestion Control Approaches for Vanets*

On the reactive side of congestion control, Khorakhun, et al. developed an algorithm, Power or Rate based congestion control, which adjusts either the power of transmission or the rate of packet generation with relation to the locally measured channel busy time ratio [19]. The channel busy time is defined as the time fraction during which the channel was sensed busy.

Depending on whether the local measurement is below or above a predefined threshold, the transmission power or generation rate is either increased or decreased by one step. In order to achieve a higher level of fairness, the authors stated that it is necessary to exchange the local measurements among neighboring vehicles, and allow an increase of the transmission power or rate only if the currently used value is below the average power or rate configuration used by the vehicle’s neighbors. Compared with proactive approaches, this reactive approach is not able to avoid congestion on the wireless channel, and supports no prioritization of different classes of messages. In addition, a simple analysis shows that the proposed algorithm is not able to prevent oscillations in the adjustment process. The issue is systematic and fundamental: since not all vehicles perform the transmit power adjustment at the same point in time, it can easily happen that the transmit power reduction at a few nodes leads to a reduced channel busy time observation from the perspective of neighboring nodes that have yet not reduced their transmit power. As a result, those nodes will possibly increase their transmit power (instead of decreasing it as well), and amplify the transmit power reduction of nodes that have already decreased their transmit power. It is obvious that some sort of additional feedback is needed to indicate the reason why the measured channel busy time has decreased or to determine who should reduce first [9].

### *C. Hybrid Congestion Control Approaches for Vanets*

A hybrid approach, attempts to combine the advantages of both proactive and reactive approaches, was proposed by Baldessari et al. in [20] which is Combined transmit power and rate congestion control (Power & Rate combined CC). Their solution consists of an improved rate control, an improved power control and a combined power and rate control algorithm, all of which use channel busy time observations to derive the number of neighbors in the surrounding area (optionally, also through an additional exchange of local vehicle density estimations). Based on the number of neighbors and a predefined channel busy time threshold, the authors then either derive a packet generation rate directly, or start via a fixed rate of packet generation and derive the maximum power of transmission which will not violate the threshold. In the latter case, the authors assume that the vehicles in the surrounding area are distributed uniformly and, typical for a proactive approach, make use of a communication model that maps carrier sense ranges to individual transmission power levels [9].

W. Zhang, A. Festag, et al. suggested that each node locally applies both the reactive and proactive approach in a distributed way in [7] which is Concepts and Framework for Congestion Control (CF for CC). So, their proposed approach can be considered as another hybrid congestion control approach. For preventing the problem of congestion channels, they also offered smart and efficient rebroadcast strategies by restricting the number of forwarded packets. Broadcasting the beacons without having any control mechanisms leads to produce a lot of redundant packets and cause the broadcast storm problem. Smart rebroadcasting scheme from [7] runs only vehicles on the same lane and located behind the accident vehicle will forward the event-driven safety message. The vehicles only forward event-driven safety message after their successful reception of this event-driven safety message from front vehicles. In real scenarios, when accident happened it's also involve other lanes. The researchers in [7] also proposed efficient rebroadcasting in their concepts and framework of congestion control. The efficient rebroadcasting will reduce the transmission rate with minimum overhead.

Besides, they proposed to use channel busy time (CBT) as metrics for network load and defined three parameters which are latency reliability and dissemination area for the network performance of safety messages.

As a result, congestion control algorithms should reduce the channel busy time in order to meet the requirements of the network performance [7].

Another hybrid congestion control approach, Adaptive Inter-vehicle Communication Control (AICC), was recently proposed in [21], where C.-L. Huang, et al. adaptively change both beacon generation rate (in a proactive way) and transmission power (in a reactive way) with the goal of reducing channel congestion, and consequently improving a vehicle's ability to accurately track the position of surrounding vehicles. Two slightly different control approaches are applied to the tuning of beacon generation rate and transmission power. Beacon generation rate is tuned based on a predicted tracking error of own position. The prediction accounts for channel unreliability, i.e., packet losses, by including the observed fraction of successfully received beacons sent by surrounding vehicles. Additionally, transmission power control is applied based on the observed channel status, more specifically, based on the channel busy time. Note that both beacon generation rate and transmission power use information locally available at the vehicles (i.e., direct observations) to control transmission parameters. As a consequence, this mechanism bears the same fundamental issue observed for [19]: without knowing the channel congestion status of the surrounding nodes, the transmission power adaptation mechanism cannot know why the channel is no longer congested and which vehicle should reduce or increase its power value first [9].

## IV. TAXONOMY OF CONGESTION CONTROL APPROACHES FOR VANETS

The proposed taxonomy is derived based on the most important characteristics of congestion control algorithms in VANETs and also from the functionality of each algorithm. This taxonomy of congestion control approaches for VANETs is illustrated in table I. According to our taxonomy, most of the proposed algorithms use transmission power and packet generation rate control for controlling the congestion in VANETs. Most of them used sensing mechanism with a pre-defined threshold for adjusting the transmit power and packet generation rate. However, by using a pre-defined threshold value optimal usage of bandwidth is not possible. This happens due to having wastage of pre-reservation of the bandwidth for some types of packets. In contrast, optimal usage of the bandwidth can be obtained by defining a dynamic carrier sense threshold instead of a fixed one. There are many types of packets with different priorities in VANETs, thus, providing a priority mechanism for different types of packets is an important task in VANETs, which are proposed in some of the reviewed algorithms in this paper. Accordingly, using the utility function, as well as smart rebroadcasting is not an optimal solution because of their complexity and high overhead.

As a result, we are able to say, one of the best congestion control algorithms for VANETs is a proactive algorithm, which uses transmission power and packet generation rate control at the same time, based on dynamic carrier sense threshold, in order to provide different priorities for different types of packets.

TABLE I. TAXONOMY OF CONGESTION CONTROL APPROACHES FOR VANETS

| Class | Approach | Ref. | Packet rate | Utility function | Power control | Access priority | Carrier sense | Smart rebroadcast |
|---|---|---|---|---|---|---|---|---|
| Proactive | VCCW | [10] | ✓ | × | × | × | × | × |
| | UBPFCC | [11] | × | ✓ | × | ✓ | × | × |
| | DPBS | [2] | × | ✓ | × | ✓ | × | × |
| | Cross-layer CC | [12] | × | × | ✓ | ✓ | × | ✓ |
| | BRR-EPA | [13] | × | × | ✓ | ✓ | × | × |
| | Application based CC | [16] | ✓ | × | ✓ | × | × | × |
| | D-FPAV | [17] | × | × | ✓ | ✓ | ✓ | × |
| Reactive | Power or Rate based CC | [19] | ✓ | × | ✓ | × | ✓ | × |
| Hybrid | Power & Rate combined CC | [20] | ✓ | × | ✓ | × | ✓ | × |
| | CF for CC | [7] | × | × | × | × | × | ✓ |
| | AICC | [21] | ✓ | × | ✓ | × | × | × |

✓ : Yes ×: No

## V. CONCLUSION AND FUTURE WORK

After In this paper, we discussed the advantages and disadvantages of the most existing congestion control algorithms in VANETs. We conclude that many of these congestion control algorithms are used proactive actions to prevent congestion in communication channel because of their characteristic which prevents congestion before it happens.

However, most of the proposed congestion control algorithms do not focusing on event-driven safety messages which are the most important type of safety messages. Furthermore, these congestion control algorithms do not address the event-driven safety messages congestion. In real situation, various reactions from drivers will generate multiple event-driven safety messages. The same priority of event-driven safety messages are also generated from different transmitters. The nodes with the same high priority packets need to be scheduled before starting the transmitting process. Thus, scheduler is needed in each node (vehicle).

In a future work, according to our findings, we will propose a proactive algorithm for congestion control in VANETs based on a combination of power transmit and generation rate control which adjust by dynamic threshold. We also plan to verify and evaluate the performance of our proposed congestion control algorithm using a network simulator such as NS-2.